\documentclass[journal,transmag]{IEEEtran}
\usepackage{booktabs}

\ifCLASSINFOpdf
  \usepackage[pdftex]{graphicx}
\else
  \usepackage[dvips]{graphicx}

\fi

\usepackage{subfigure}

\usepackage{subfiles}
\usepackage[colorinlistoftodos]{todonotes}

\begin{document}
%
\title{Predicting Award Winning Research Papers at Publication Time}


\author{\IEEEauthorblockN{Vella Riccardo, 
Vitaletti Andrea, Silvestri Fabrizio}}

%



\IEEEtitleabstractindextext{%
\begin{abstract}
%
%
%
In recent years, many studies have been focusing on predicting the scientific impact of research papers. Most of these predictions are based on citations count or rely on features obtainable only from already published papers. In this study, we predict the likelihood for a research paper of winning an award only relying on information available at publication time.  For each paper, we build the citation subgraph induced from its bibliography. We initially consider some features of this subgraph, such as the density and the global clustering coefficient, to make our prediction. Then, we mix this information with textual features, extracted from the abstract and the title, to obtain a more accurate final prediction. We made our experiments considering the ArnetMiner citation graph, while the  ground truth on award-winning papers has been obtained from a collection of best paper awards from 32 computer science conferences. In our experiment, we obtained an encouraging F1 score of $0.694$. Remarkably, The high recall and the low false negatives rate, show how the model performs very well at identifying papers that will not win an award. This behavior can help researchers in getting a first evaluation of their work at publication time. Lastly, we made some first experiments on interpretability. Our results highlight some interesting patterns both in topological and textual features.
\end{abstract}

}

\maketitle

\IEEEdisplaynontitleabstractindextext

\newcommand{\specialcell}[2][c]{%
  \begin{tabular}[#1]{@{}l@{}}#2\end{tabular}}

%
\IEEEpeerreviewmaketitle
\section{Introduction} \label{Introduction}
\IEEEPARstart{T}{he} 
escalating growth of published papers, estimated to increase exponentially with an average doubling period of 15 years \cite{fortunato2018scienceofscience}, poses a challenge in navigating through this huge amount of information. One natural question arises: can we identify impactful research?
\newline
Previous research on this topic~\cite{yu2014citationimpactprediction, burrell2003predicting, bruns2016research} use features that are available only after a paper has been published (for example, the trend of the number of citations). This does not provide a useful measure for scholars to assess their work's quality promptly.

Numerous studies have relied on citation counts for their predictions, yet the effectiveness of these approaches has been extensively debated. The field of research significantly affects citation numbers, as highlighted by \cite{radicchi2012citationfairness} on citation fairness. Furthermore, certain types of citations, such as self-citations and obligatory citations \cite{hemmat2015mandatoryselfcitations} (on mandatory self-citations), offer less value and are challenging to identify. These issues point to the conclusion that citations do not hold uniform value.

\vspace{5pt}
\textit{Contribution of the paper.} To address the limitations identified earlier, this study introduces a novel methodology for forecasting the future influence of a paper within the academic community. A pertinent question that arises is how to effectively define what constitutes an impactful paper, specifically establishing a solid foundation for our predictions. In this context, this research categorizes an impactful paper as one that has been recognized with an award. Thus, our goal is to assess the probability of a paper receiving an award, relying solely on the data available at the time of its publication. Based on this premise, we will demonstrate that the citation network accessible at the point of publication provides a sufficiently reliable basis to predict a paper's future success. 
Our experimental results show that considering both topological and textual features of the citation network, we can achieve an encouraging $0.694$ F1 score with a simple model. Due to the nature of the problem, the model is naturally good at predicting papers that will not win an award. We also assess the interpretability of our proposed model by performing some initial experiments showing that award-winning papers have textual and topological features that are separated from the same features of non-award-winning contributions.

\section{Related Work} \label{Related Work}
%
%
%
There is a rich literature concerning the objective of predicting the impact of research papers. The works in this field can be reviewed under two scopes: the features, and method, used for prediction and the objective of the prediction. 
\newline
Many types of features have been used in the prediction. Different studies make use of features which are not available at publication time, like early citations \cite{yu2014citationimpactprediction}\cite{burrell2003predicting}\cite{bruns2016research}, or even altmetrics \cite{eysenbach2011cantweetspredict}\cite{timilsina2016towards}.
Others make use of information extracted from the graph and available at publication time. Sarigöl et al. \cite{sarigol2014coauthorship} find a correlation between coautorship networks and highly cited papers. Livne et al. \cite{livne2013textgraph} as well as Pobiedina and Ichise \cite{pobiedina2016citationlink} use graph mining to extract different features and predict the citation count. Klimek et al. \cite{klimek2016successful} predict citation impact after using text to build a measure of document centrality. 
\newline
When it comes to the objective of the prediction, a large amount of studies focuses on predicting citation counts (Ibáñez et al. \cite{ibanez2009bioinformatics}, Cummings and Nassar \cite{cummings2020citationgnn}, Abrishami and Aliakbary \cite{abrishami2019citationdeep}, Livne et al. \cite{livne2013textgraph}, Pobiedina and Ichise \cite{pobiedina2016citationlink}) or highly cited papers (Newman \cite{newman2014highlycited}, McNamara et al. \cite{mcnamara2013highimpact}, Sarigöl et al. \cite{sarigol2014coauthorship}). Other studies focus on predicting different impact measures like the PageRank (Sayyadi et al. \cite{sayyadi2009futurerank}) or the h-index (Dong et al. \cite{dong2015hindex}\cite{dong2016hindex}).

A very small fraction of the literature includes awards in the prediction as a features or an objective. Yang et al. \cite{yang2011award} predict award winners by extracting temporal features from the citation graph, while Liu and Huang \cite{liu2019towards} make use of the co-authorship network. Fiala and Tutoky \cite{fiala2017pagerankaward} evaluate the results of citation count and PageRank predictions in the task of predicting awards. 

\section{Approach} \label{Approach}
\begin{figure}[h]
    \centering
    \includegraphics[width=0.48\textwidth]{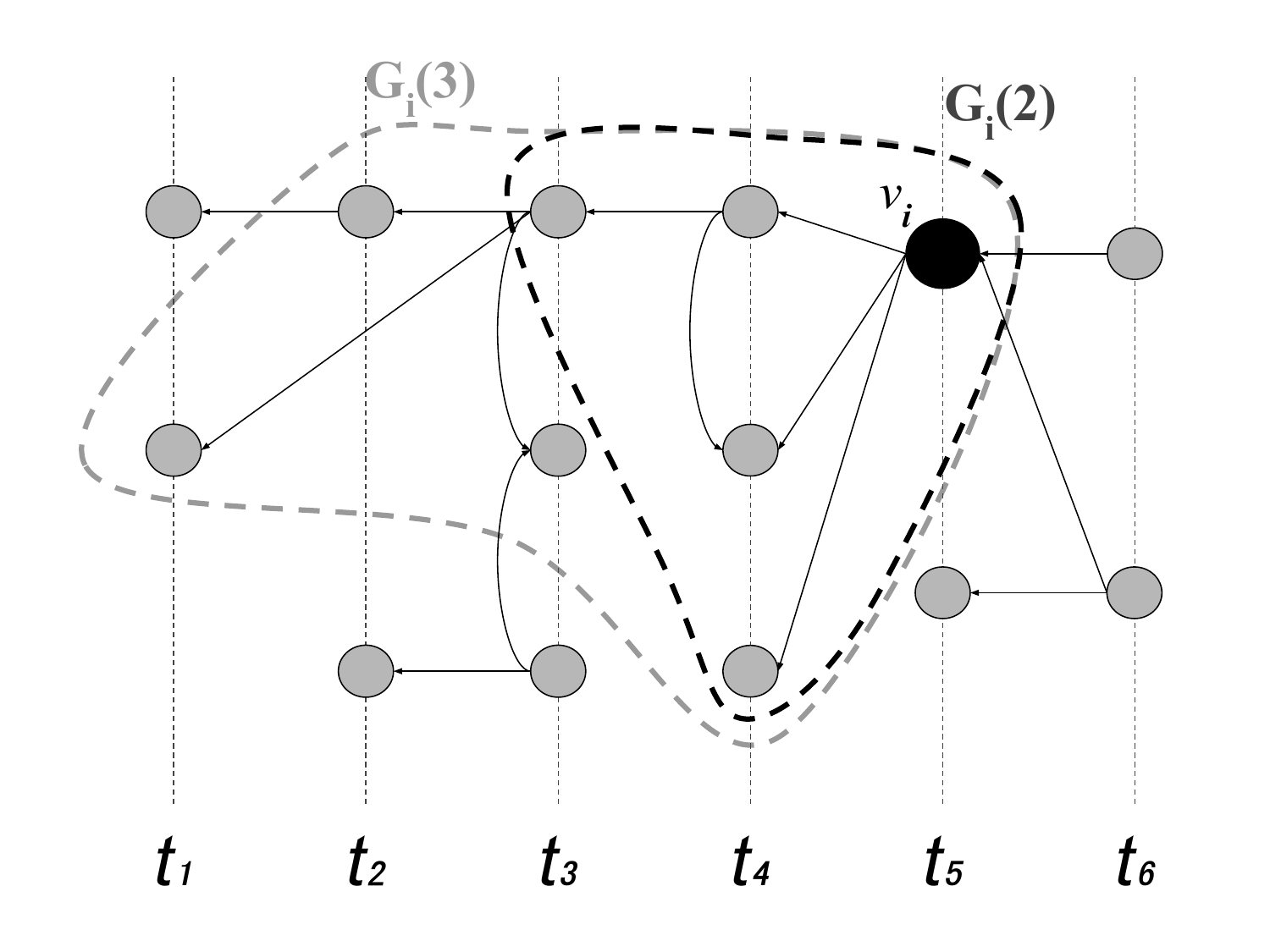}
    \caption{An example of a citation graph where every node is located at a certain time $t$. The example shows how the choice of the $\Delta$ parameter affects the resulting subgraph $G_i(\Delta)$. } 
    \label{fig:award_graph_example}
\end{figure}

Research papers are represented as the nodes of a graph structure which we call citation graph or citation network. In such a directed and unweighted graph $G = (V, E)$, the nodes are the research papers $V = \{v_1, v_2, ..., v_n\}$, where $n$ is the total number of papers and $E$ is the set of edges such that  $e_{i, j} = (v_i, v_j) \in E$ if the research paper $v_i$ cites $v_j$. 
We define the subgraph $G_i(\Delta) = \{V_i(\Delta), E_i(\Delta)\} \subseteq G$ (see Figure \ref{fig:award_graph_example}), where $V_i(\Delta) \subseteq V$ is the set of nodes formed by $v_i$ and all the nodes in $G$ reachable from $v_i$ in $\Delta$ steps, i.e.
$$
\forall \hspace{2pt} v_j \in V; \hspace{5pt} v_j \in V_i(\Delta) \iff d(v_i, v_j) \leq \Delta
$$
with $d(v_i, v_j)$ representing the shortest path length between $v_i$ and $v_j$. $E_i(\Delta)$ is the set of edges that we can walk from $v_i$ to any $v_j$ in $\Delta$ steps.
We say that a subgraph $G_i(\Delta)$ is a winning graph if the research paper $v_i$ has won an award. 
\newline
\textit{Considering any subgraph $G_i(\Delta)$, can we predict the probability of this graph being a winning graph? What would the accuracy of this prediction be?}

\subsection{Topological Features} \label{Topological Features}
For each subgraph $G_i(\Delta)$, we compute a set of five simple topological features $\chi_{1,i}$: \emph{Average Out-Degree},\emph{Diameter}, \emph{Network Density}, \emph{Global Clustering Coefficient (or Transitivity)}, \emph{Average Local Clustering Coefficient}.
For a definition of these measures and details on computation we refer to Boccaletti et al. \cite{boccaletti2006complexnetworks} and to the NetworkX library \cite{NetworkX}.

These features are classified with a multi-layer perceptron, with one hidden layer. We focus on this simple classifier for two main purposes: a) a non-linear classifier provides better performance (see Section \ref{Experiments}); b) we aim at building the simplest possible model to ease the investigation on the interpretability of our results.



\subsection{Textual Features} \label{Textual Features}
Apart from the topological features, we obtain another feature vector $\chi_{2,i}$ as the output of TinyBERT Transformer \cite{jiao2019tinybert}, given as input the abstract or, if the abstract is not present, the title of the paper $v_i$, to obtain a sentence-level embedding. 
\newline
Like for the other features, the classification is obtained through a second multi-layer perceptron. This is a common fine tuning practice \cite{devlin2019bert}.

\subsection{Final Dataset}
The citation graph is obtained from the ArnetMiner dataset \cite{arnetminer}, a collection of publications in Computer Science, and merged with award data from a collection of best paper award for 32 computer science conferences \cite{huang2023awards}. After the initial processing, it contains more than 3 million papers, of which 838 are award winners.

To speed up the computation, from this graph, we extract a set of $n=10000$ subgraphs $G_i(\Delta)$, with $i = 1,...,n$, as in the previously described methodology (Section \ref{Approach}). The extracted set is obtained from all the available award winners plus a set of non-winners, randomly sampled from the graph. 
\newline
Since the graph mostly comprises papers that are topologically very far from award winners, for fairness, we perform a weighted random sampling. To do so, we group non-winner papers in subsets $S_a \subset V$ based on their distance from the closest award. More specifically, considering a non-winner paper $v_j$ and its closest award winner $v_i$, $v_j \in S_a \iff d(v_j, v_i) = a$, being $d(v_j, v_i)$ the shortest path length from $v_j$ to $v_i$. After grouping, we obtain the weight of $v_i$, with distance $a$ from the closest award, as $w_i = \frac{n}{||S_a||}$. This way, we can sample with a uniform distribution of distances from the closest award.

The final dataset can be written as $D = \{\chi_1, \chi_2, y\}$, with $\chi_1 = \{\chi_{1,1},...,\chi_{1,n}\}$ being the set of all topological features, $\chi_2 = \{\chi_{2,1},...,\chi_{2,n}\}$ being the set of all textual features and $y = \{y_1,...,y_n\}$ being the set of all output labels, where an output label is defined as 
\[
y_i = \left\{ \begin{array}{rcl}
 1 & \mbox{if} & v_i \mbox{ is an award winner} \\ 
 0 & \mbox{otherwise}
\end{array}\right.
\]

\subsection{Final Mixed Model} \label{Final Model}
As described above in this section, we obtain a first classification probability $\hat{y}_1 = \Gamma_1(\chi_1)$, using a multi-layer perceptron, with a single hidden layer, as the classifier $\Gamma_1$. A second classification probability is obtained from $\hat{y}_2 = \Gamma_2(\chi_2)$, in the same fashion.
\newline
The final prediction is found as
\[
\hat{y}_{1,2} = \Gamma_{1,2}(\{\Gamma_1(\chi_1), \Gamma_2(\chi_2)\})
\]
Here $\Gamma_{1,2}$ is a third perceptron with two input nodes, which are the prediction probabilities from the other models, and returns a final prediction probability, which can then be rounded to obtain the final classification.

\section{Results} \label{Results}
In this section, we show and discuss the results. All the results were obtained using the dataset built as described in Section \ref{Approach}, with $\Delta = 2$. Since this is a binary classification problem on an unbalanced dataset, for evaluation, we use the following metrics: F1-score, ROC AUC, Precision-Recall AUC.
   
\begin{table}[ht]
  \caption{Prediction Performance}
  \label{table:results}
  \begin{tabular}{lccc}
   \toprule
    Features&ROC&Preciosion-Recall &F1\\
   &AUC & AUC & Macro\\
    \midrule
    Topological & 0.811 & 0.284 & 0.558\\
   Textual & 0.756 & 0.248 & 0.635\\
    Topological + Textual & 0.915 & 0.518 & 0.694\\
 \bottomrule
\end{tabular}
\end{table}


%

\begin{figure}[ht]
\centering
\subfigure[\label{fig:pr_curve}]{\includegraphics[width=1.65in]{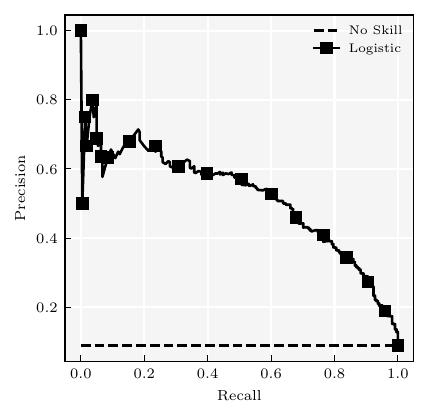}} 
\hfill
\subfigure[\label{fig:year_dist}]{\includegraphics[width=1.65in]{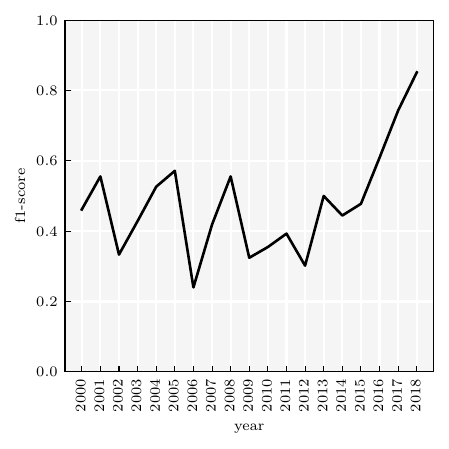}}
\caption{Precision-Recall \ref{fig:pr_curve} curve of the final mixed model and the model's F1-score, on evaluation, visualized over the year of the predicted papers \ref{fig:year_dist}.}
\label{fig:roc_pr}
\end{figure}

%
The low Precision-Recall AUC (see Figure \ref{fig:pr_curve}) is explained by a large percentage of false positives. This effect is intrinsic to the nature of the task of predicting awards and it is expected. As an example, not all successful papers win awards, even though they might have all the characteristics to do so. The high recall and the low false negatives rate, show how the model performs very well at identifying papers that will not win an award, even though it is insecure on the papers that will actually win. This behaviour can make this model an interesting tool for researchers to evaluate their work. 
   
The F1-score is similar for all the awards, namely the model is not particularly biased towards a specific research topic. Moreover, Figure \ref{fig:year_dist} highlights how the model is better at classifying more recent papers. There can be two causes for this effect: a) older data is less complete and therefore less reliable; b) the factors related to winning an award are dependent on time.
If there is a time-dependent prediction, the model shows better performance on newer data simply because the dataset contains more recent papers and it would be possible to further enhance the overall performance by adopting a solution that considers time during training.
\begin{figure}[ht]
\centering
\subfigure[\label{fig:phi_score}]{\includegraphics[width=1.65in]{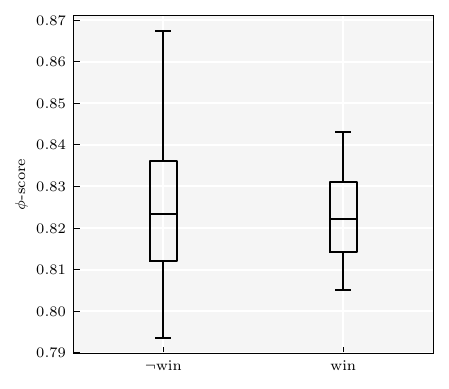}} 
\hfill
\subfigure[\label{fig:theta_score}]{\includegraphics[width=1.65in]{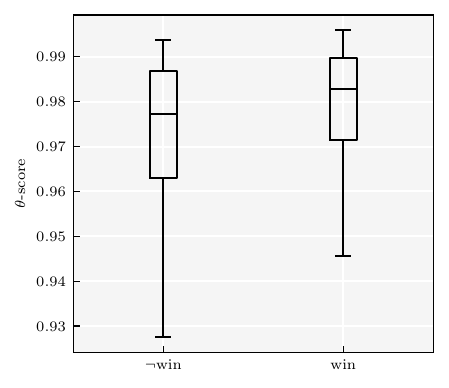}}
\caption{A comparison of the distributions of all $\phi$-scores \ref{fig:phi_score} and $\theta$-scores \ref{fig:theta_score} for winners and non-winners. Especially from the $\phi$-score experiment, it is clear that the scores of the winners lie in a specific range, while the distribution of the non-winners, has similar mean but a greater variance.}
\label{fig:phi_theta_scores}
\end{figure}

\section{First Experiments on Interpretability} \label{Experiments} 
In this section, we experiment on the model and on the data, trying to provide a first interpretation of the obtained results. 
   
\begin{figure*}[ht]
    \centering
    \includegraphics[width=0.6\textwidth]{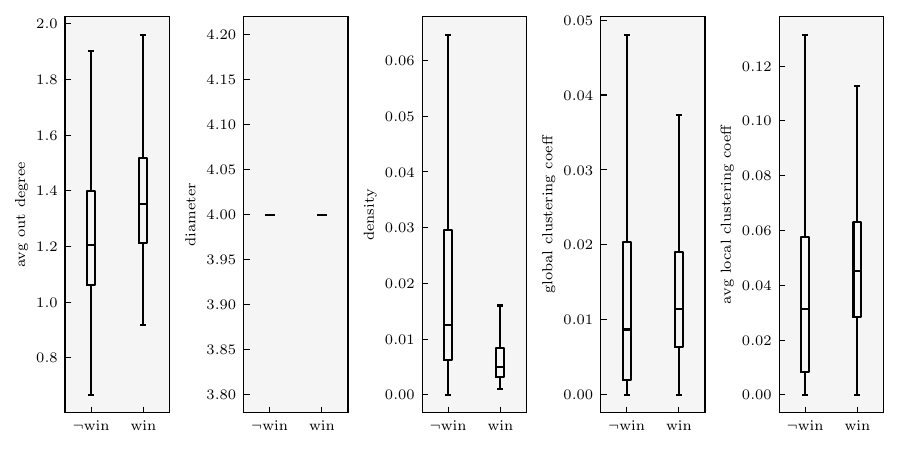}
    \caption{A comparison of the distributions of the topological features for winners and non-winners, in the form of box plots. The horizontal axis represents the true label $y$.}
    \label{fig:features_visualized}
\end{figure*}

To better understand the model's behavior in classifying the topological features, we build a simplified version of the topological features based model (described in Section \ref{Topological Features}). This version $\bar{\Gamma}$, is a simple perceptron, trained to classify features $\chi_1$, which achieves a \textit{F1-score} of $0.492$ (the score of our final topological model is $0.558$). We then analyze the weights of the model to get insights on the most relevant features. The \emph{Network Density ($-65.9$)} is by far the most significant weight, followed by the \emph{Global Clustering Coefficient ($-17.5$)} and the \emph{Average Local Clustering Coefficient ($7.0$)}. All the remaining features have weights close to $1.0$. These weights, together with the distributions shown in Figure \ref{fig:features_visualized}, prove that network density and global clustering coefficient are the most influential features. Indeed, for those features in particular, winners reside in a bounded area, while non-winners are more spread out. These results hold also when features are normalized. 
\newline
A strongly negative feature for the network density signifies that the model tends to predict a paper $v_i$ as a winner, if the corresponding subgraph $G_i(\Delta)$ has low connectivity, i.e., in our case, if the papers cited by $v_i$ cite each other "less". 
\newline
This interpretation, that can be made also for the global clustering coefficient, highlights an interesting behavior. A paper that cites papers that already cite each other will probably not bring a new perspective on the field and will probably be less impactful. Impactful papers, more likely, may be the ones which can represent a bridge between different topics, merging them in an innovative way that will lead to new solutions.

In this scenario, we would also expect textual features to reflect this behaviour, meaning that the textual features of articles cited by winning papers should be less similar than usual. In the following experiments we build measures that can be representative of this behaviour and see if they fit our interpretation.

\subsection{Similarity of Textual Features}
For each paper $v_k$, with subgraph $G_k(\Delta) = \{V_k, E_k\}$, we consider the set of papers $\bar{V_k} = V_k - \{v_k\}$ and define $X_k$ as the set of textual features for each paper in $\bar{V_k}$. We want to compute a score that is informative of the textual similarity between all papers $\bar{V_k}$. 
\newline
We define the text features similarity between two papers $v_i, v_j$, with textual features $x_i, x_j$, as the cosine similarity
$$
\cos(x_i,x_j) = \frac{x_i \cdot x_j}{\left\| x_i\right\|_{2}\left\| x_j\right\|_{2}}
$$
\newline
From this definition, we construct a score of similarity $\phi_k$, for a paper $v_k$, between the set of all cited papers $\bar{V_k}$, with features $X_k$, as the average cosine similarity between all pairs of textual features in $X_k$
$$
\phi_k = \frac{\sum_{x_i,x_j \in X_k; i \neq j} \cos(x_i, x_j)}{n(n-1)}
$$
where $n$ is the number of cited papers in $\bar{V_k}$.
$\phi_k \in [-1, 1]$ and to an higher $\phi$-score it corresponds a higher similarity of textual features $X_k$.

In Figure \ref{fig:phi_score}, we see how the $\phi$-score of the winners fall within a specific range, while the scores of the non-winners are more dispersed. This result extends our previous interpretation. Articles cited by winning papers are not just less similar, but they lie in a specific range of similarity.

\subsection{Clustering Textual Features}
In the second experiment we define and compute a new score, based on the number of clusters in which we can group all textual features from a paper $v_k$.

This time, we define $n_c$ as the number of clusters obtained running the DBSCAN algorithm \cite{ester1996dbscan} on the set of features $X_k$ and define the $\theta$-score as $ \theta_k = {n_c} / n $,
with $\theta_k \in [0, 1]$. A higher $\theta$-score corresponds to a lower tendency for features to form clusters and, therefore, a greater spreading, overall, of textual features $X_k$.

In this case, results, shown in Figure \ref{fig:theta_score}, are less clear, but there is still some difference between winners and non-winners. In fact, once again, we can notice a smaller variance for winners, as well as a smaller tendency to group in clusters.

\section{Conclusion and Future Work} \label{Conclusion}
%
%
%
The problem of predicting the impact of a research paper has been addressed in many different ways, often centered around the citations count. These solutions can be effective, but they are usually not applicable before the paper is published, making it less useful as a tool for the researchers that may want to evaluate their work at writing time.
\newline
We propose a novel approach to solve the problem of predicting the impact for a research paper, by modeling this impact as the chance of winning an award in the future. Our content based approach is able to compute a time independent prediction, even before the publication of the paper, and is built by first focusing on topological features, then on textual features and finally by combining the models resulting from these two sets of features. 
\newline
We build a dataset specifically for this purpose, starting from the citation network built from a large collection of papers in the field of Computer Science and apply it in the training of our model. We then evaluate the results obtained with our method and simplify them to obtain an interpretation, finding that some properties of the neighbourhood of a paper are descriptive for our task. In particular the density, global clustering coefficient and average local clustering coefficient. Lastly we highlight some properties of the winners neighbourhood through experiments.

In our work, we build the problem around a clear and unambiguous ground truth, that is awards winning. We address the problem while aiming at simplicity, obtaining encouraging results and showing the feasibility of an accurate prediction. We show that this results are interpretable and that using a mix of different features leads to better results.
\newline
Further research, that can reach even better results, may include: considering the whole dataset instead of a sample; building a more powerful and sophisticated classifier (maybe making use of Graph Neural Networks); conducting further research on interpreting the behaviours described in this work.

\ifCLASSOPTIONcaptionsoff
  \newpage
\fi



\bibliographystyle{IEEEtran}
%
\bibliography{references} 

\end{document}